\documentclass{article}
\usepackage[pdftex]{graphicx}
\usepackage{amsmath}
\usepackage{amssymb}
\usepackage{float}
\usepackage{url}
\usepackage{authblk}

\begin{document}
\title{A Neural Network Alternative to Non-Negative Audio Models}

\author{Paris Smaragdis$^{\sharp,\flat}$ \qquad Shrikant Venkataramani$^{\sharp}$\thanks{Partial funding for this work was provided by the National Science Foundation under award number 1453104}} 
\affil{$^{\sharp}$University of Illinois at Urbana-Champaign\\
    $^{\flat}$Adobe Research}

\maketitle

\begin{abstract}
We present a neural network that can act as an equivalent to a Non-Negative Matrix Factorization (NMF), and further show how it can be used to perform supervised source separation. Due to the extensibility of this approach we show how we can achieve better source separation performance as compared to NMF-based methods, and propose a variety of derivative architectures that can be used for further improvements.
\end{abstract}

\section{Introduction}
During the last decade, we have seen an increasing use of Non-Negative Models for audio source separation \cite{tsp1,tsp2}. In this paper we describe an alternative computational approach for such models that makes use of neural networks. The reason for this approach is to take advantage of the multiple conveniences of neural network models that allow us to design non-negative model variants that are overcomplete, multi-layered, arbitrarily non-linear, have temporal structure, can address non-linear mixtures, etc. Additionally, this approach allows us to effortlessly implement new architectures due to the wealth of automatic differentiation tools available for this purpose. As we will show in this paper, using a neural network approach also allows us to obtain significantly improved results.

In this paper we will address two of the main issues that need to be resolved to implement a non-negative model using a neural network; the calculation of a non-negative basis representation from an audio signal, and the calculation of a non-negative latent state from an audio signal. Using these two steps we can easily replicate most of the existing literature in non-negative models. In the remainder of this paper we will introduce a process for these two calculations and then show how they compare with a traditional non-negative audio model in separation tasks.

\section{Non-Negative Autoencoders}
\subsection{A non-negative autoencoder architecture}
The well-known $K$-rank Non-Negative Matrix Factorization (NMF) model as introduced in \cite{nmf} is defined as:
\begin{equation}
    \mathbf{X} \approx \mathbf{W} \cdot \mathbf{H}
\label{nmf}
\end{equation}
where $\mathbf{X} \in \mathbb{R}^{M\times N}_{\geq 0}$ is a non-negative input matrix to approximate, and $\mathbf{W} \in \mathbb{R}^{M\times K}_{\geq 0}$, $\mathbf{H} \in \mathbb{R}^{K\times N}_{\geq 0}$ are the two non-negative factors, commonly referred to as bases and activations respectively. The set $\mathbb{R}^{M\times N}_{\geq 0}$ is that of real, non-negative matrices of size $M \times N$. This factorization has been the core element for many source separation methods in the last few years \cite{tsp1,tsp2}.

Let us now reinterpret the NMF model as a linear autoencoder. The obvious formulation is:
\begin{equation}
\begin{aligned}
    &\footnotesize{\textsf{1\textsuperscript{st} layer:}}& \mathbf{H} &= \mathbf{W}^\ddagger \cdot \mathbf{X} \\
    &\footnotesize{\textsf{2\textsuperscript{nd} layer:}}& \hat{\mathbf{X}} &= \mathbf{W} \cdot \mathbf{H}
\end{aligned}
\label{nae1}
\end{equation}
in which we enforce the constraint that $\mathbf{W},\mathbf{H} \geq 0$. The non-negative matrices $\mathbf{W}$ and $\mathbf{H}$ would correspond to their namesakes in the NMF model, whereas the matrix $\mathbf{W}^\ddagger$ would be some form of a pseudoinverse of $\mathbf{W}$ that produces a non-negative $\mathbf{H}$. The output $\hat{\mathbf{X}}$ is the model's approximation to the input $\mathbf{X}$. In autoencoder terminology, the first layer weights $\mathbf{W}^\ddagger$ are referred to as the \textit{encoder} (which produces a code representing the input), and the upper layer weights $\mathbf{W}$ are referred to as the \textit{decoder} (which uses the code to reconstruct the input). Although this representation would be functionally equivalent to NMF, it would not exhibit any specific advantage and is more complicated and burdensome to implement. Instead we use a slightly different formulation that, as we will show later on, has more interpretative power and is more in line with common neural network designs. Consider the Non-Negative Autoencoder (NAE) model:
\begin{equation}
\begin{aligned}
    &\footnotesize{\textsf{1\textsuperscript{st} layer:}}& \mathbf{H} &= g\left(\mathbf{W^\ddagger} \cdot \mathbf{X} \right) \\
    &\footnotesize{\textsf{2\textsuperscript{nd} layer:}}& \hat{\mathbf{X}} &= g\left(\mathbf{W} \cdot \mathbf{H} \right)
\end{aligned}
\label{nae2}
\end{equation}
where $g : \mathbb{R}^{M\times N} \mapsto \mathbb{R}^{M\times N}_{\geq0}$, i.e. an element-wise function that produces non-negative outputs. Well-known examples of such functions in the neural network literature include the rectified linear unit: $g(x)=\max(x,0)$, the softplus: $g(x)=\log( 1 + e^x)$ or even the absolute value function $g(x)=|x|$. By applying such an activation function we ensure that our latent representation $\mathbf{H}$ and that the approximation $\hat{\mathbf{X}}$ are both non-negative. There is no guarantee that the matrices $\mathbf{W}^\ddagger$ and $\mathbf{W}$ will be non-negative, but that is not a necessary constraint as long as the output and latent state are. 

There are of course many ways to estimate the two weight matrices $\mathbf{W}^\ddagger$ and $\mathbf{W}$, but for the remainder of this paper we will use the following approach. The entire input $\mathbf{X}$ will be used as a single batch and the parameter updating will be estimated using the RProp algorithm \cite{rprop}. For the activation function we will use the softplus function \cite{softplus}. We will use the cost function from \cite{nmf}:
\begin{equation}
D(\mathbf{X},\hat{\mathbf{X}}) = \sum_{i,j} \left(\mathbf{X}_{i,j} \left[\log( \mathbf{X}_{i,j}) - \log( \hat{\mathbf{X}}_{i,j})\right] - \mathbf{X}_{i,j} + \hat{\mathbf{X}}_{i,j}\right)
\label{cost}
\end{equation}
where the subscripts $i,j$ acts as indices on the matrix they are applied on.
\begin{figure*}[ht]
\centering
\includegraphics[width=\textwidth]{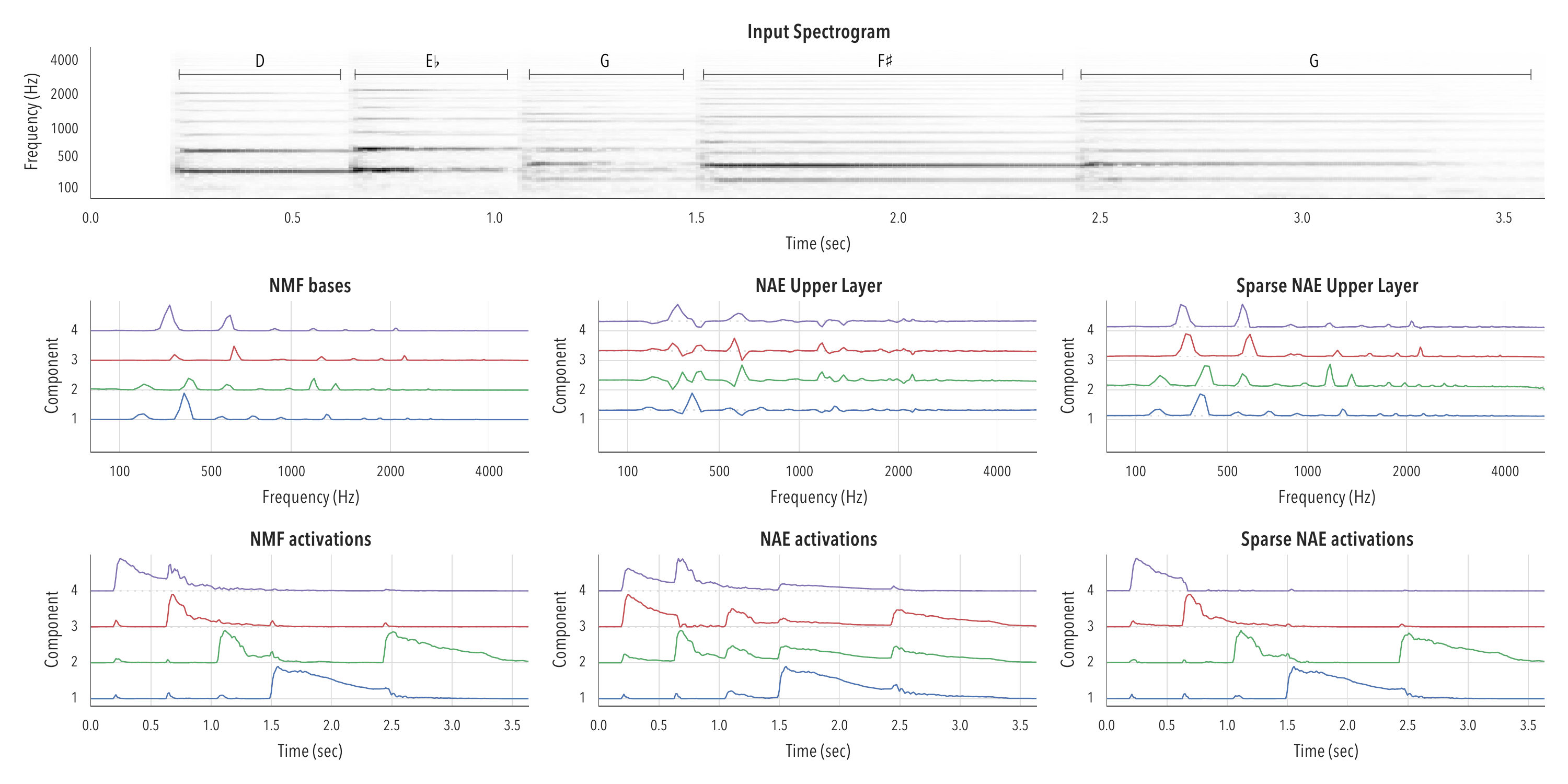}
\caption{A comparison between NMF and NAEs for component discovery in spectrograms. The input spectrogram is shown in top plot, and consists of five notes, as labelled. The left plots show the coding discovered by NMF. The NMF bases and activations correctly identify the spectral shape and activation of the four pitches. The middle plots show the results for an NAE, with the upper layer's matrix rows shown at the top plot and the first layer outputs shown at the bottom plot. Although they approximate the input well, they are not as perceptually meaningful. At the right we see the results from a sparse NAE. Asking for a sparse first layer output results in an encoding that's equivalent to NMF.}
\label{plot1}
\end{figure*}
\subsection{Learning a non-negative model}
\label{learnmodels}
In the context of audio processing, NMF is often used to decompose the magnitude of time-frequency distributions (e.g. the input matrix $\mathbf{X}$ is often a magnitude spectrogram). To illustrate the differences of this model with NMF when using such inputs, consider the following example from \cite{pariswaspaa} shown in figure \ref{plot1}. This is a snippet of a piano performance, with the note sequence \{D, E$^\flat$, G, F$^\sharp$, G\}, which is four distinct pitches (each having a different spectrum), with the pitch G repeated twice making this a five-note sequence. This structure can be clearly seen in the spectrogram in figure \ref{plot1}.

As shown in \cite{pariswaspaa}, we can extract that information by modeling the magnitude spectrogram using equation \ref{nmf} with a rank of 4. We set $\mathbf{X}$ as the input spectrogram, and upon convergence we obtain $\mathbf{W}$ and $\mathbf{H}$, which will respectively hold the model's four \textit{bases} and their corresponding four \textit{activations}. Upon visual examination we see that the columns of $\mathbf{W}$ resemble the spectra of the four notes in the input, whereas the rows of $\mathbf{H}$ activate that spectra at the right points in time to construct the input. These are shown in the middle and bottom left plots in figure \ref{plot1}.

We now turn to the problem of estimating a similar encoding using a neural network. Using the model in equation \ref{nae2} we obtain the matrices $\mathbf{W}$ and $\mathbf{H}$. These essentially represent an NMF model, albeit with $\mathbf{W}$ potentially having negative values, and having the non-negativity of the output being enforced by the nonlinearity $g(\cdot)$ (which is set to softplus in this case). This model learns a good representation, but it isn't as intuitive as the NMF model. We see that the spectral bases take on negative values which will result in some cross-cancellation being used for the approximation, thereby obfuscating the component-wise additive structure of the model.

One way to resolve the problem of basis cross-cancellations is to use regularization. We see that in this case multiple bases are activated simultaneously, forcing each unique spectrum in the input to be represented by multiple bases at a time. This is a very redundant coding of the input resulting in an unnecessarily busy activation pattern. By adding a sparsity regularizer on $\mathbf{H}$ we can obtain a more efficient coding of the input and minimize activation redundancy. We do so by extending the cost function in equation \ref{cost} by:
\begin{equation}
\mathcal{L} = D(\mathbf{X},\hat{\mathbf{X}}) + \lambda ||\mathbf{H}||_1
\end{equation}
We repeat the above experiment using this new cost function and report the result in the right plots of figure \ref{plot1}. As is clearly evident, this model learns a representation which is qualitatively equivalent to NMF (in fact it is slightly more efficient due to the regularization).

\subsection{Learning a latent representation given a model}
\label{fit}
Having learned a model for a sound, we now turn to the problem of extracting the activations $\mathbf{H}$ for an input sound if the bases $\mathbf{W}$ are already known. This is a crucial process for non-negative audio models since it allows us to explain new signals given already learned models. In the case of NMF this is a very straightforward operation; the estimation is the same as learning the full model but we keep the matrix $\mathbf{W}$ fixed. In the case of the NAE model this operation isn't as obvious. Ideally we would expect to pass a new input through the first layer of a trained NAE and obtain an estimate of $\mathbf{H}$ for that sound, but this is not a reliable estimator when using mixtures of sounds. Fortunately the solution to this problem isn't complicated and is a reinterpretation of the neural network training process.

Consider the model in equation \ref{nae2}. For the task at hand we will be given an input spectrogram $\mathbf{X}$ and a learned model $\mathbf{W}$ and we would have to estimate an $\mathbf{H}$ such that $\hat{\mathbf{X}} \approx \mathbf{X}$. This essentially becomes a single-layer non-linear network:
\begin{equation}
\hat{\mathbf{X}} = g( \mathbf{W} \cdot \mathbf{H})
\end{equation}
In usual neural network problems we would be given a target $\mathbf{X}$ with corresponding inputs $\mathbf{H}$ and would be expected to learn the model weights $\mathbf{W}$. What we have to solve in this case is a complementary problem, where we are given the targets and weights but we need to estimate the inputs. This is of course simply the original problem with the dot product operands swapped, and we can easily solve it using simple gradient backpropagation\footnote{by transposition: $\hat{\mathbf{X}}^\top = g( (\mathbf{W}\cdot\mathbf{H})^\top) = g( \mathbf{H}^\top \cdot \mathbf{W}^\top)$ this becomes the same as a generic training problem where we can estimate $\mathbf{H}^\top$ by pretending it is a weight matrix} (or any other variants of neural network learning).
\subsection{Extensions}
Since we now make use of a neural network framework we can easily implement extensions of this model. The most obvious case would be the one of a multilayered (or deep) network, as opposed to the \textit{shallow} model presented above. In this case we implement the NAE as:
\begin{equation}
\begin{aligned}
\mathbf{Y}_0 & = \mathbf{X} \\
\mathbf{Y}_i &= g( \mathbf{W}_i \cdot \mathbf{Y}_{i-1}), i = 1,2,...,2L \\
\mathbf{H} &= \mathbf{Y}_{L}\\
\hat{\mathbf{X}} &= \mathbf{Y}_{2L}
\end{aligned}
\end{equation}
where we use $2L$ layers overall, and we ensure that the layer sizes are symmetric about the middle, i.e. that if $\mathbf{W}_i \in \mathbb{R}^{M \times N}$ then $\mathbf{W}_{2L+1-i} \in \mathbb{R}^{N \times M}$. The output of the $L$'th layer $\mathbf{H}$ will be the latent representation. This model effectively uses the first $L$ layers as an encoder to produce a latent representation, and then uses the upper $L$ layers as a decoder and produces an approximation of the output $\hat{\mathbf{X}}$. Just as before, we minimize the previously used cost function between the network's output $\hat{\mathbf{X}}$ and input $\mathbf{X}$, and train using the same methods as before. This kind of model will allow us to use more complex representations of the input with a richer dictionary, which would be impossible to simulate with NMF.

Additionally, we can use more exotic layer types, and implement each layer using a recurrent neural network (RNN) and its variants \cite{LSTM,GRU} to make use of temporal context, or use convolutional layers \cite{conv} to make use of time/frequency context, or any of the many flavors of neural network layers that are available today. In this paper we will limit our discussion to the two models explicitly described above, but it is very easy to implement any other layer type for additional modeling power.

\section{Supervised Separation}
We now turn our attention to the problem of supervised separation \cite{parisica}. In this setting, NMF is often used to learn an a priori model of the types of sounds, and then once presented with a new mixture containing such sounds, the learned models are used to decompose that mixture into the contribution of each source. We therefore have two steps or processing, one being the training of the source models, and the other being the fitting of these models on a mixture sound. We have addressed both of these problems in the previous sections, for the source separation problem we need to add a couple more details discussed below.

The first step for this source separation process is to learn models of the sounds we expect to encounter. We can simply do that independently for each sound type using the methodology shown in section \ref{learnmodels}. The only information that we need to retain would be the decoders of the learned NAEs, which will be used to compose an approximation of the input mixture.

Once the decoders are obtained, the next step is to use them simultaneously to explain a mixture containing sounds relating to them. To do that we will combine them using the following setup:
\begin{equation}
\begin{aligned}
\hat{\mathbf{X}}_1 &= g( \mathbf{W}_1 \cdot \mathbf{H}_1) \\
\hat{\mathbf{X}}_2 &= g( \mathbf{W}_2 \cdot \mathbf{H}_2) \\
\hat{\mathbf{X}} &= \hat{\mathbf{X}}_1 + \hat{\mathbf{X}}_2
\end{aligned}
\label{mnae}
\end{equation}
where $\mathbf{W}_i$ are the already obtained decoders, one for each sound class. We use each decoder to approximate an output $\hat{\mathbf{X}}_i$ and then we sum these outputs to produce an approximation of the input mixture $\mathbf{X}$. The only parameters we can adjust to achieve this approximation are $\mathbf{H}_i$, the latent representations of the two models. Conceptually this problem isn't much different than the problem in section \ref{fit} and is easy to solve using standard methods. One optional change we can make at this point is to add one more regularizer to discourage models being active simultaneously in a redundant way. We do so by setting the cost function to:
\begin{equation}
\mathcal{L} = D(\mathbf{X},\hat{\mathbf{X}}) + \lambda \sum_k ||\mathbf{H}_k||_1
\end{equation}
This regularizer usually results in a modest improvement, but is by no means necessary.

In the case of a multilayer NAE (or any other layer type), the above equations need to be extended to include the entire decoder of the pretrained models. To explain a mixture we would only need to estimate just the inputs to the first layer of these decoders, and then sum their outputs.
\section{Experiments}
We now present some experiments separating speech mixtures to compare the NAE approach to a traditional NMF method. We composed 32 0dB mixtures of two random TIMIT speakers \cite{timit}, and for each speaker's 10 sentences we use 9 to train a speaker model and 1 to use in the mixture. For preprocessing we take the magnitude spectrogram of the mixture using a 512pt DFT, applying a square-root Hann window, and a hop size of 25\%. The magnitude spectra of the training data and the mixture are being used as inputs to an NMF or NAE estimator. To reconstruct the extracted sources from $\hat{\mathbf{X}}_i$ we use:
\begin{equation}
s_i(t) = \operatorname{STFT}^{-1}\left( \frac{\hat{\mathbf{X}}_{i}}{\sum_{j} \hat{\mathbf{X}}_{j}} \odot \mathbf{X} \odot e^{ \mathrm{i} \mathbf{\Phi}}\right)
\end{equation}
where $\mathbf{X}_i$ is the estimated magnitude spectrogram of the $i$'th source, and $\mathbf{\Phi}$ is a matrix containing the phase of the input mixture. The operator $\odot$ denotes element-wise multiplication, and  $\operatorname{STFT}^{-1}(\cdot)$ is the inverse spectrogram, which produces a waveform from a complex-valued spectrogram. We run two experiments to measure the performance of this approach. We measured the success of separation using the median BSS\_EVAL metrics \cite{bsseval} and STOI index \cite{stoi}.

\begin{figure*}[ht]
\centering
\includegraphics[width=\textwidth]{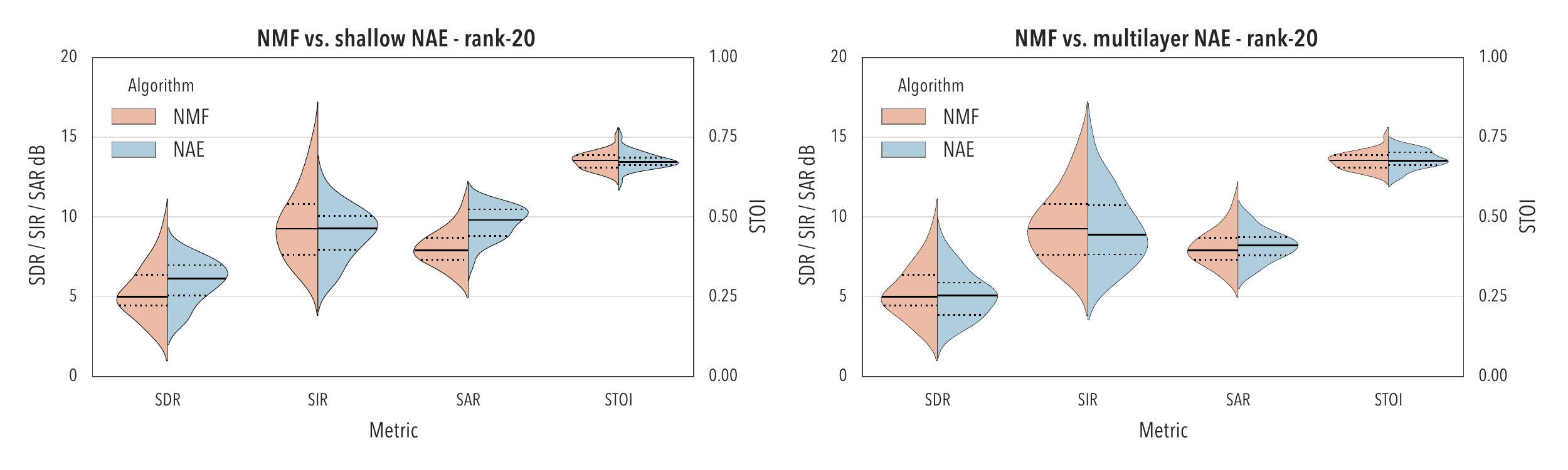}
\includegraphics[width=\textwidth]{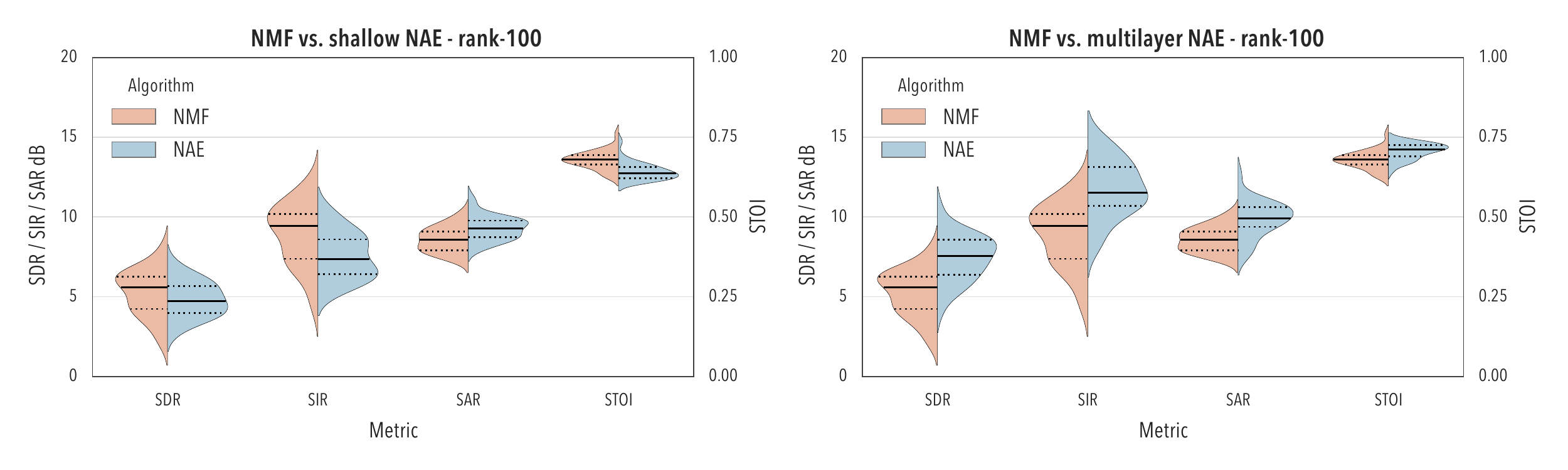}
\caption{Comparison of source separation performance on speech/speech mixtures between NMF and NAEs. The left-facing (pink) distributions are of NMF separation results, whereas the right-facing (blue) distributions are of NAE results. The thick solid line in each distribution shows the median value over all experiments, and the dashed lines delimit the corresponding interquartile range. The top plots compare the results between rank-20 NMF with a 20 unit NAE (left), and a rank-20 NMF with a four-layer NAE ($L=2$) with 20 units in each layer (right). The bottom plots show the same type of comparison for models with rank 100.}
\label{violin}
\end{figure*}
\begin{figure}[]
\centering
\includegraphics[width=.45\textwidth]{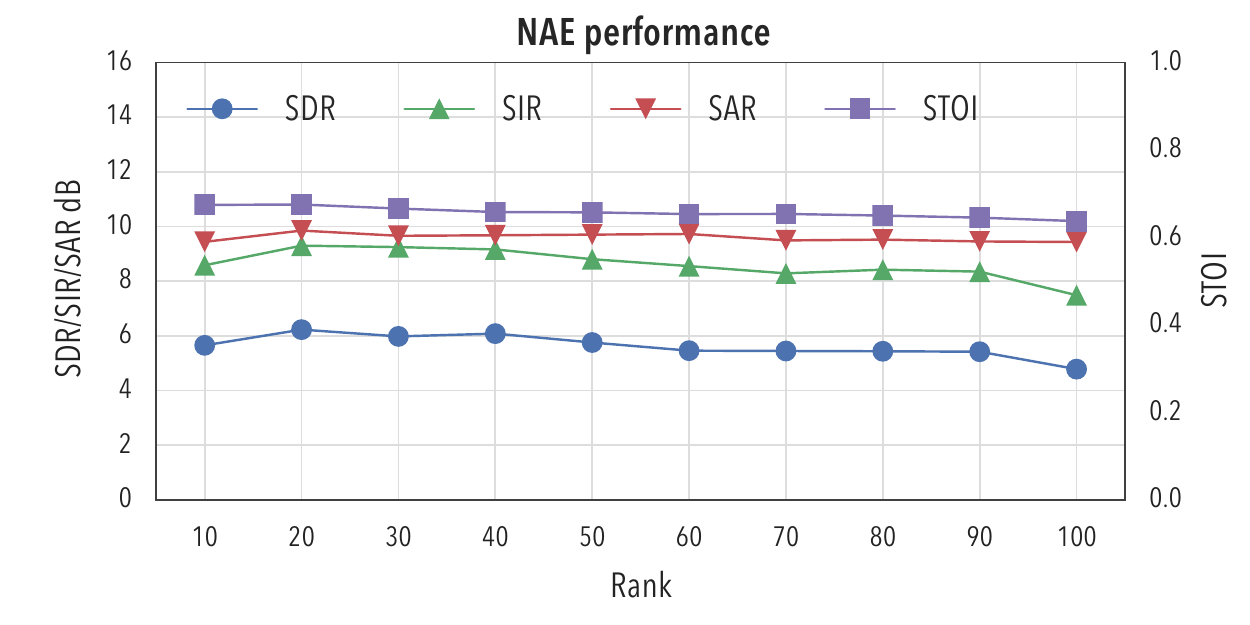}
\includegraphics[width=.45\textwidth]{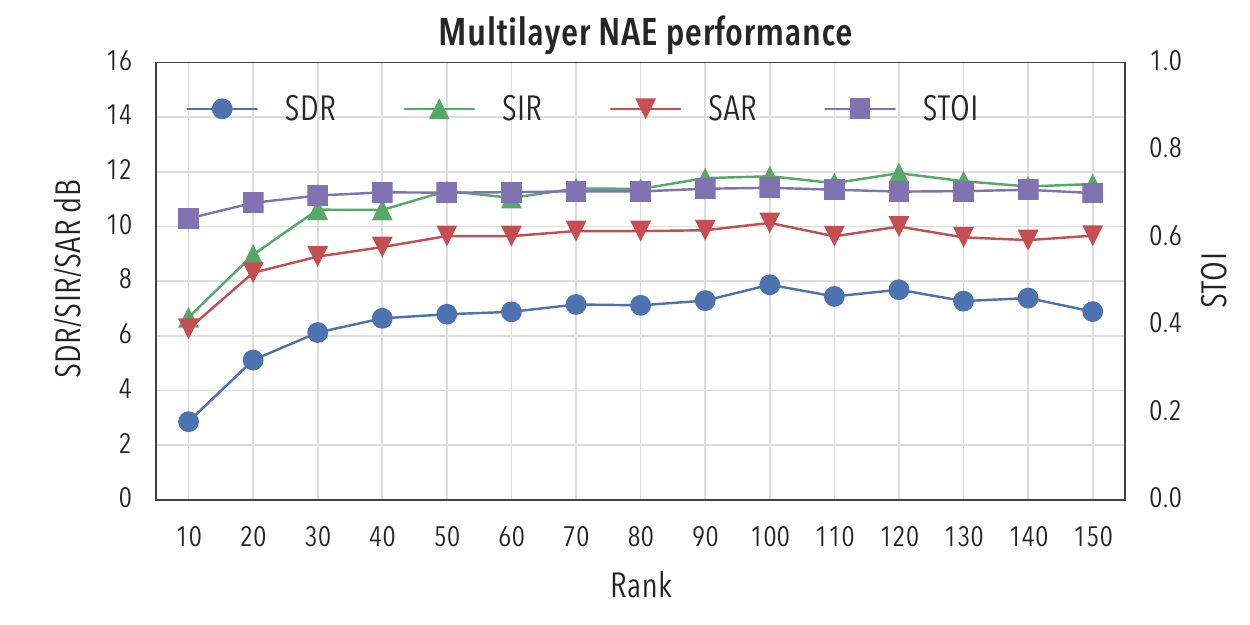}
\caption{Performance on speech/speech mixtures of shallow (left) and multilayer (right) NAEs with varying number of components. For speech/noise mixtures the results are generally 2 to 4 dB higher.}
\label{comps}
\end{figure}

The first experiment compared  the ability of the NAE model to resolve mixtures when using various layer sizes. The results of are shown in figure ~\ref{comps}. We see that for a shallow NAE (equation~\ref{nae2}) performance peaks around 20 components (roughly the same behavior as with NMF separators). For a multilayer NAE (equation \ref{mnae}, $L=2$ and all layers being the same size), we see that performance increases as we add more components, and peaks at 100.

We also compared a basic NMF separator with a shallow and a multilayer NAE ($L=2$) of the same size (figure~\ref{violin}). In general, we see that the shallow NAE performs roughly equivalently to NMF separators, albeit with worsening performance when using higher rank decompositions (which is expected since as shown before, shallow NAE performance degrades at large ranks). For the multilayered NAE, we see that it matches NMF performance with a rank of 20, but performs significantly better for a rank of 100 (again this is expected from the results in figure~\ref{comps}). Note that for the large NAE the interquartile range of the results is above the interquartile range of NMF, implying a consistently better performance.
\section{Conclusions}
In this paper we presented an alternative approach to applying non-negative models for source separation. We show that this approach results in significantly improved performance, and that it lends itself to a wealth of model extensions that would be difficult to implement using the traditional NMF methodology.
\newpage
\bibliographystyle{IEEEbib.bst}
\bibliography{nae}

\end{document}